\documentclass[a4paper,12pt]{article}
\usepackage{amsmath,amssymb,amsthm,latexsym,graphicx,float}
\pdfoutput=1

\newtheorem{prop}{Proposition}

\numberwithin{equation}{section}

\title{\sc How Big Should a Stress Shock Be?}
\author{{David G. Maher}\footnote{DavidGMaher@yahoo.com.au}}

\begin{document}

\maketitle

\newcommand{\D}{\mathbf{D}}
\newcommand{\R}{\mathbb{R}}
\newcommand{\C}{\mathbb{C}}
\newcommand{\E}{\mathbb{E}}
\newcommand{\Z}{\mathbb{Z}}
\newcommand{\N}{\mathbb{N}}
\newcommand{\T}{\mathbb{T}}
\newcommand{\F}{\mathcal{F}}
\newcommand{\A}{\mathcal{A}}
\newcommand{\x}{\mathbf{x}}
\newcommand{\y}{\mathbf{y}}
\newcommand{\sgn}{\mathrm{sgn} \,}
\newcommand{\dd}{\, \mathrm{d}}
\newcommand{\Tr}{\, \mathrm{Tr}}

{\abstract Stress shocks are often calculated as multiples of the standard deviation of a history set.  This paper investigates how many standard deviations are required to guarantee that this shock exceeds any observation within the history set, given the additional constraint of kurtosis.  The results of this analysis are then used to validate the shocks produced by some stress test models, in particular that of Brace-Lauer-Rado.\\
\indent A secondary application of our results is to investigate three known extensions of Chebyshev's Inequality where the kurtosis is known.  It is found that our results give a tighter bound than the well-known inequalities.\\

{\bf Keywords:} Stress model validation, kurtosis, Chebyshev's Inequality.}

\section{Introduction}

A common way to determine stress shock sizes is as a multiple, $k$, of the daily standard deviation, $\sigma$, of the risk factor.  $\sigma$ is typically calibrated on history set of daily rate changes of a suitable length, $N$. The multiple, $k$, is sometimes referred to as a {\it tail factor}.  This can be computed from the inverse of the cumulative distribution function.  For example, if the returns were modelled on the normal distribution, then the tail factor for a 1-in-1,000,000 day shock would be 4.75.\\

Stress models are rarely published as 1) they are proprietary, and 2) there is no regulatory requirement to do so.  As a result, to our knowledge there is no literature on the performance or validation of these models.  Indeed, stress models cannot be effectively backtested as they are designed to produce shocks large enough so as not to have backtesting exceptions.\\

The following stress models have been communicated to the author:
\begin{itemize}
  \item Several banks assume that returns follow  a Student-t distribution with low degrees of freedom (c.f. \cite{M}).  For example, 3 degrees of freedom (the lowest degree for which the distribution has a finite variance) produces a tail factor of $103.3$.  The length of the history set varies between banks.
  \item One Australian bank uses a tail factor of 7, with an additional add-on for less liquid assets.  The  history set used for calibration is the VaR history set of 2 years.
  \item One Canadian Bank and one Australian bank use the stress model of Brace-Lauer-Rado \cite{BLR}, which is primarily parameterised by a given value of kurtosis to produce a tail factor. The length of the history set varies between each of these banks.\\
\end{itemize}

{\bf This paper investigates the following question:} Given a stress shock of size $k \sigma$, can we be assured that this shock exceeds any observed shock in the historical data?  After all, if a shock was found in the last year or two to be larger than the derived stress shock, then this effectively invalidates the stress model.\\

The approach of this paper to this problem is to extend the results of Samuelson \cite{S} (see also \cite{J} and the references therein).  Samuelson shows that no single value can lie more than $\sqrt{N-1}$ deviations from the mean by examining the endpoint example where one observation is equal to 1, with the remaining $N-1$ observations set to zero.  The main result of this paper is to impose conditions on the kurtosis of this type of endpoint distribution (section 2), which is then applied to validate the stress test models presented above, in particular that of Brace-Lauer-Rado \cite{BLR} (section 3).  This is a novel approach, which could be considered the first to analyse stress models with a falsifiable test.\\

Following the outline of Samuelson \cite{S}, we then use this endpoint distribution to examine three extensions of Chebyshev's Inequality where the kurtosis is known.  These are a version of Chebyshev's Inequality with higher moments of even order, Zelen's Inequality \cite{Z}, and Bhattacharyya's Inequality \cite{Ba}.  Similarly to Samuelson's results, we give a tighter bound for three known extensions of Chebyshev's inequality in the finite case.\\


\section{Incorporating Kurtosis into Samuelson's distribution}

In \cite{S}, Samuelson shows that no single value can lie more than $\sqrt{N-1}$ deviations from the mean.  This is achieved by constructing a finite point distribution where one observation is equal to 1, with the remaining $N-1$ observations set to zero.  This distribution is unrealistic as a model for extreme moves of asset prices.  It would be equivalent to the market never ever moving except for a large jump one day in 10,000 (or even more), and that our estimate of the volatility is driven entirely by the size of that jump.  The reality is that asset prices do move, with varying and clustered volatility, and with more frequent smaller jumps too.\\

To incorporate these features into Samuelson's distribution, we will place a restriction on the kurtosis.  Focussing on kurtosis is ideal for the validation of a stress model as it does not impose any other assumptions about the distribution, only a general metric of how heavy-tailed the distribution is.\\

Thus, let us consider distributions which have low kurtosis, plus one outlier.  The bi-modal distribution with points at $\pm 1$, which has a kurtosis of 1, is the obvious candidate.  This distribution, for a fixed value of kurtosis, enables the value of $(X_1 - \bar{X})/\sigma$ to be large.\\

To rigorously prove that a distribution, which for a fixed value of kurtosis, maximises the value of $(X_1 - \bar{X})/\sigma$, is a complex optimisation problem, analogous to that posed in \cite{S}.  To give some certainty this distribution cannot be improved upon, it was compared to three other distributions in the Appendix: neither produced a larger value for $(X_1 - \bar{X})/\sigma$ than the bi-modal.  But moreover, $(X_1 - \bar{X})/\sigma$ did not differ very much for any choice of distribution, giving credibility to our choice of kurtosis being a good restriction to make, as it allows for asset prices to move - and move in different ways - with some impact on the volatility aside from the one large shock.\\

So to paraphrase \cite{S}: How deviant can one be when the distribution has a restriction of kurtosis imposed?  Consider the bi-modal distribution with one extreme point, which we call $\Phi_N$:
$$
\Phi_N = (X_1, X_2, X_3 , X_4, X_5, \dots , X_N) = (a, b, -b, \dots , b, -b)
$$
with $a$, $b$ and $c$ are normalised such that $\E(X) = 0$, $\E(X^2) = 1$, and $\E(X^4) = \kappa$.  Thus, $a$ is the maximum possible number of standard deviations away from the mean.  We prove the following:\\

\begin{prop}\label{FormulaFora} Suppose $N \geq 5$ is odd.  The value $a = a(N,\kappa)$ is given by

$$
a = a(N,\kappa) = \sqrt{ -\frac{N-1}{N+1} + \sqrt{ \biggl( \frac{N-1}{N+1} \biggr)^2 - G(N,\kappa)}}
$$

where
$$
G(N,\kappa) = \frac{N(N-1)^2 - (N-1)^3 \kappa}{(N+1)(N-3)}
$$

Furthermore, $a \sim [N (\kappa - 1)]^{1/4}$ for large $N$.\\

\end{prop}

{\bf Proof:}  Suppose that the mean of the non-normalised distribution is $\frac ay$.  To ease computation, assume that $N$ is odd.  Subtracting $\frac ay$ from every observation, except $X_1$, gives
\begin{align*}
\E(X) = \frac 1N \sum_{i=1}^N X_i & = \frac aN + \frac{N-1}{2N} (-b - \frac ay) + \frac{N-1}{2N} (b - \frac ay) \\
& = \frac aN - \frac{N-1}{N}.\frac ay \\
\Rightarrow y = N-1 &
\end{align*}

From the second moment:
\begin{align*}
\E(X^2) = \frac 1N \sum_{i=1}^N X_i^2 & = \frac{a^2}{N} + \frac{N-1}{2N} (-b - \frac{a}{N-1})^2 + \frac{N-1}{2N} (b - \frac{a}{N-1})^2 \\
& = \frac{a^2}{(N-1)} + \frac{N-1}{N} b^2 \\
& \Rightarrow b^2 = \frac{N}{N-1} - a^2 \frac{N}{(N-1)^2}
\end{align*}

Since $\E(X) = 0$ and $\E(X^2) = 1$, the kurtosis $\kappa$ is simply $\E(X^4)$:
\begin{align*}
\kappa = \E(X^4) = \frac 1N \sum_{i=1}^N X_i^4 & = \frac{a^4}{N} + \frac{N-1}{2N} (-b - \frac{a}{N-1})^4 + \frac{N-1}{2N} (b - \frac{a}{N-1})^4 \\
& = \frac{a^4}{N} + \frac{N-1}{2N} (2b^4 + 12\frac{b^2 a^2}{(N-1)^2}  + 2\frac{a^4}{(N-1)^4})
\end{align*}

Substituting $b^2 = \frac{N}{N-1} -a^2 \frac{N}{(N-1)^2}$:
\begin{align*}
& \kappa = \frac{a^4}{N} + \frac{N-1}{N} \biggl( \Bigl(\frac{N}{N-1} - a^2 \frac{N}{(N-1)^2} \Bigr)^2 + 6\frac{\Bigl(\frac{N}{N-1} -a^2 \frac{N}{(N-1)^2} \Bigr) a^2}{(N-1)^2}  + \frac{a^4}{(N-1)^4} \biggr) \\
&= \frac{a^4}{N} + \frac{N-1}{N} \biggl( \Bigl( \frac{N^2}{(N-1)^2} - \frac{2N^2a^2}{(N-1)^3} + \frac{N^2a^4}{(N-1)^4} \Bigr) + \frac{6Na^2}{(N-1)^3} - \frac{6Na^4}{(N-1)^4} + \frac{a^4}{(N-1)^4} \biggr) \\
&= \frac{a^4}{N} + \frac{N(N-1)}{(N-1)^2} + \frac{(6-2N) a^2}{(N-1)^2} + \frac{(N^2-6N+1)a^4}{N(N-1)^3} \\
&= \frac{N(N-1)}{(N-1)^2} + \frac{6-2N}{(N-1)^2} a^2 + \frac{N^2-6N+1 + (N-1)^3}{N(N-1)^3} a^4 \\
&= \frac{N(N-1)}{(N-1)^2} + \frac{-2(N-3)}{(N-1)^2} a^2 + \frac{(N+1)(N-3)}{(N-1)^3} a^4 \\
\end{align*}

Rearranging, and multiplying by $\frac{(N-1)^3}{(N+1)(N-3)}$ gives the following quadratic in $a^2$:

\begin{equation}\label{aaa}
0 = (a^2)^2 - 2\frac{N-1}{N+1} (a^2) + G(N,\kappa)
\end{equation}

where
$$
G(N,\kappa) = \frac{N(N-1)^2 - (N-1)^3 \kappa}{(N+1)(N-3)}
$$

The value of $a$ as a function of $\kappa$ and $N$ is given by the quadratic formula, then taking the square root of the positive case.  This gives this first statement of the proposition.\\

For the second statement, observe that letting $N$ become large yields
$$
a = a(N,\kappa) = \sqrt{ -1 + \sqrt{ 1 + N(\kappa - 1)}} \sim [N (\kappa - 1)]^{1/4}
$$

for large $N$.  $\phantom{abcde} \square$\\

{\bf Remark:} $a$ (the number of standard deviations above the mean) grows with a leading term of $N^{1/2}$ in Samuelson's distribution, but our distribution grows much slowly in $N^{1/4}$.  The fact that $a$ also grows proportionately to $[N (\kappa - 1)]^{1/4}$ for large $N$ will be important regarding Chebyshev's Inequality in section 4.\\

Values of $a$ for a given $N$, and various values of kurtosis, are tabulated below:\\

\begin{table}[H]
\begin{tabular}{| c | c || c | c | c | c |}
\cline{1-6}
& & \multicolumn{4}{ | c | }{a = (Max-mean)/StDev} \\
\hline 		
N-1 &	sqrt(N-1)	& kurtosis = 7	&   kurtosis = 10 &	kurtosis = 13 &	kurtosis = 16 \\
\hline
250  &  15.811  &  6.296  &  6.952  &  7.46  &  7.881  \\
500  &  22.361  &  7.464  &  8.247  &  8.853  &  9.355  \\
1,000  &  31.623  &  8.855  &  9.789  &  10.511  &  11.109  \\
10,000  &  100.000  &  15.682  &  17.349  &  18.638  &  19.705  \\
100,000  &  316.22  &  27.849  &  30.817  &  33.113  &  35.011  \\
1,000,000  &  1,000.000  &  49.502  &  54.781  &  58.865  &  62.241  \\
833,208  &  912.802  &  47.296  &  52.339  &  56.241  &  59.467  \\

\hline
\end{tabular}
\caption{\label{table2}$a = (Max-mean)/StDev$ for given $N$ and kurtosis.}
\end{table}

\medskip
\medskip

Thus, if a bank were to target a AA rating, it would need to consider its survivability over a 833,209 day ($\approx 3333$ year) history set.  For the unconstrained case, a tail factor of approximately 913 standard deviations would be required.  For a distribution with the choices of kurtosis presented here, a tail factor of 47 to 60 standard deviations would be required.  This result in a stress shock of a large magnitude, we would consider it far more reasonable.\\

There may be other considerations for the size of the history set considered  -  not least the fact that a 3,333 year history set of daily prices is impossible to obtain!  For example, the Basel Committee on Banking Supervision has recently recommended as part of its ``Fundamental review of the trading book" (see \cite{B}) that the Internal Model be calibrated on a history set of 10 years.  This choice of history set length results in a tail factor of 11 to 14 standard deviations with our above choices of kurtosis presented here.  This is still a reasonably sized stress shock, though obviously not as large.\\

In fact, the above results can be used to construct a stress shocks, without any distributional assumptions beyond kurtosis.  For example, $N = 833,209$ is best for targeting a AA rating, $N = 10,000$ is suitable for the consideration of once-in-generation shocks, and $N = 3,000$ for shocks that occur in a typical business cycle (say).\\

In the next section a validation of the shocks produced by a stress model that emphasises kurtosis is performed by comparing the shocks given to the results above.

\medskip
\medskip



\section{Validation of Stress Models}

In this section the above result are applied to validate the stress shock that are produced by the three stress models outlined in the Introduction: 1) Stress shocks based on the Student-t distribution, 2) a tail factor of 7,  calibrated on a history set of 2 years, and 3) the stress model of Brace-Lauer-Rado \cite{BLR}.

\subsection{Student-t Distribution}

The Student-t distribution has a finite standard deviation only when the degrees of freedom are above 2, and finite kurtosis only when the degrees of freedom are above 4.  Tabulated below are the tail factors for the Student-t distribution with degrees of freedom 3, 4, 5 and 6.  Against these are the values of $a(N, \kappa)$ when $\kappa = $ 6 and 3, corresponding to the degrees of freedom cases 5 and 6.\\

Only when the degree of freedom is 3 can it be assured that the stress shock will exceed the maximum value when the constraint of kurtosis is in place.  But as this distribution does not have a finite value of kurtosis, the comparison cannot be made.

\begin{table}[h]
\begin{tabular}{| c || c | c | c | c || c | c |}
\cline{1-7}
& \multicolumn{4}{ | c | }{Tail Factor} \\
\hline 		
Deg. freedom & 3 & 4 & 5 & 6 & & \\
\hline
$N \, \backslash \, \kappa$ & N/A & N/A & 6 & 3 & $a (N, \kappa = 6)$ & $a (N, \kappa = 3)$ \\
\hline
250  &  6.322  &  4.908  &  4.262  &  3.898  &  6.023  &  4.828  \\
500  &  8.053  &  5.951  &  5.030  &  4.524  &  7.138  &  5.709  \\
1,000  &  10.215  &  7.173  &  5.893  &  5.208  &  8.466  &  6.760  \\
10,000  &  22.204  &  13.034  &  9.678  &  8.025  &  14.987  &  11.934  \\
100,000  &  47.928  &  23.332  &  15.547  &  12.032  &  26.610  &  21.171  \\
1,000,000  &  103.299  &  41.578  &  24.771  &  17.83  &  47.298  &  37.619  \\
\hline
\end{tabular}
\caption{\label{table3a}Tail Factors for the Student-t distribution, versus $a(N, \kappa)$.}
\end{table}

\medskip
\medskip

\subsection{Tail factor of 7}

The tail factor of 7 used by one Australian bank would appear to be quite inadequate.  According to the values in Table 1, a risk factor with a kurtosis of 7 could produce a value that exceeds the stress shock when calibrated on a two year history set.

\subsection{Brace-Lauer-Rado Model}

In \cite{BLR}, the authors present a model for determining stress test shocks for a risk factor $X_t$, based on the stochastic volatility models of Scott \cite{Sc} and Wiggins \cite{W}:

$$
X_t = \exp Y_t \phantom{abcde} dY_t = \exp \tfrac12 V_t dW_t^{(1)}
$$

$$
dV_t = -gV_t dt + h dW_t^{(2)} \phantom{abcde} \langle dW_t^{(1)}, dW_t^{(2)} \rangle = \rho \, dt
$$

\medskip

The model has three parameters: $\rho$, the skewness, $g$, which controls the rate of mean-reversion in volatility back to the Normal over time, and $h$, which is connected to the kurtosis.  This model has a number of desirable features for stress tests.  For example, it possesses fat tails, but it is relatively straightforward to calculate quantiles for liquidity holding periods.\\

The instantaneous kurtosis of $Y_t$ is given by:
\begin{equation}
\kappa := \frac{\E[dY_t^4]}{\E[dY_t^2]^2} = 3 \exp \frac{h^2}{2g}
\end{equation}

As values of $h$ lead to unique values of $\kappa$ for a given $g$, the model can be re-cast in terms of $\rho$, $g$, and $\kappa$.  The authors then present the tail factors targeting a AA rating, ie, a survival probability of 0.9997 various values of $g^{-1}$ (ie, the expected time of mean reversion), and $\rho = 0.5$.  In fact, the authors present these tail factors for various liquidity holding periods, but here we just consider the 1 day holding period:\\

\medskip

\begin{table}[h]
\begin{tabular}{| c || c | c | c | c |}
\cline{1-5}
& \multicolumn{4}{ | c | }{Tail Factor} \\
\hline 		
$g^{-1}$	& kurtosis = 7	&   kurtosis = 10 &	kurtosis = 13 &	kurtosis = 16 \\
\hline
1m  &  13.648  &  17.485  &  20.445  &  22.873  \\
2m  &  13.397  &  17.148  &  20.041  &  22.412  \\
3m  &  13.278  &  16.986  &  19.846  &  22.190  \\
4m  &  13.204  &  16.886  &  19.726  &  22.053  \\
5m  &  13.153  &  16.817  &  19.642  &  21.958  \\
6m  &  13.115  &  16.765  &  19.579  &  21.886  \\
\hline
\end{tabular}
\caption{\label{table3}Tail Factors of the Brace-Lauer-Rado Stress Model.}
\end{table}

\medskip
\medskip

Comparing these tail factors to the values of $a = (Max-mean)/StDev$ in Table 1, the tail factors given by the model for the case of kurtosis = 7 and $g^{-1} = 6m$ could be violated in a history set of 5,000 observations  -  that is, 20 years of data (assuming 250 business days to a year).  For the remaining cases of kurtosis = 10, 13 and 16, this is 9,000, 12,250, and 15,250, respectively. That is, 36, 49, and 61 years of data, respectively.\\

It could be said that these values are a little low, given that our earlier results showed that for targeting the survivability of a AA rated institution.  However, only in few cases can a daily history set be obtained that is of this length, so the existence of an observation in the history set that exceeds the stress shock will be very rare.

\medskip

\section{Chebyshev's Inequality with Higher moments}

As noted in Samuelson \cite{S}, the construction of these distributions is closely connected to Chebyshev's inequality.  In this section we show that our endpoint distribution $\Phi_N$ gives a tighter bound than three extensions of Chebyshev's Inequality where the kurtosis is known  -  to our knowledge, these are the only three that consider the use of kurtosis.  These are a version of Chebyshev's Inequality with higher moments of even order, Zelen's Inequality \cite{Z}, and Bhattacharyya's Inequality \cite{Ba}.  It is found that Bhattacharyya's Inequality is not sharp.\\

Chebyshev's inequality is given by:
$$
P(|X - \bar{X}| \geq t \sigma) \leq \frac{1}{t^2}
$$

In a finite world of $N$ observations, setting $t = \sqrt{N}$ gives the endpoint:
$$
P(|X - \bar{X}| \geq \sqrt{N} \sigma) \leq \frac{1}{N}
$$
which Samuelson's example marginally betters to
$$
P(|X - \bar{X}| \geq \sqrt{N-1} \sigma) \leq \frac{1}{N}
$$

To apply our results concerning $\Phi_N$, observe the Bi-modal distribution with one extreme point $a$ (normalised such that $\E[X] = 0$ and $\E[X^2] = 1$ and $\E[X^4] = \kappa$) yields the inequality
$$
P(|X - \bar{X}| \geq a) \leq \frac{1}{N}
$$
so if an extended Chebyshev's inequality gives a threshold greater than $a$, or a probability less than $\tfrac{1}{N}$, then our finite point distribution $\Phi_N$ has been shown to give a tighter bound.  We now consider the three extensions Chebyshev's inequality mentioned above.

\medskip
\medskip

\subsection{Higher moments of even order}

For any integer $k>0$
$$
P \Bigl(|X - \bar{X}| \geq t \E[(X - \E[X])^{2k}]^{1/2k}\Bigr) \leq \frac{1}{t^{2k}}
$$
which is a straightforward consequence of Markov's inequality.  See \cite{MU}, for example.  Setting $t = N^{1/4}$ and $k = 2$ gives the endpoint:
$$
P \Bigl(|X - \bar{X}| \geq (\kappa N)^{1/4} \Bigr) \leq \frac{1}{N}
$$

Values of $(\kappa N)^{1/4}$ are tabulated below.  Note that these are all greater than the values of $a$ given in Table 1 in Section 2, and greater than the limit of $((\kappa-1) N)^{1/4}$.  That is, in a finite world of $N$ observations, $\Phi$ (our distribution conditioned on kurtosis) improves on this instance of Chebyshev's inequality.\\

\begin{table}[h]
\begin{tabular}{| c || c | c | c | c |}
\cline{1-5}
& \multicolumn{4}{ | c | }{$(\kappa N)^{1/4}$} \\
\hline 		
N-1 &	 kurtosis = 7	&   kurtosis = 10 &	kurtosis = 13 &	kurtosis = 16 \\
\hline
250  &    6.468 	& 7.071 	& 7.550 	& 7.953  \\
500  &    7.692 	& 8.409 	& 8.979 	& 9.457  \\
1,000  &  9.147 	& 10.000 	& 10.678 	& 11.247  \\
10,000  &    16.266 	& 17.783 	& 18.988 	& 20.000  \\
100,000  &    28.925 	& 31.623 	& 33.766 	& 35.566  \\
1,000,000  &    51.437 	& 56.234 	& 60.046 	& 63.246  \\
\hline
\end{tabular}
\caption{\label{table4}Values of $(\kappa N)^{1/4}$.}
\end{table}

\medskip
\medskip

\subsection{Zelen's Inequality}

For the case of where the third and fourth moments are known, formulae have been given by Zelen \cite{Z}:
$$
P(|X - \bar{X}| \geq t \sigma) \leq \biggl[ 1 + t^2 + \frac{(t^2 - t \theta_3 - 1)^2}{\theta_4 - \theta^2_3 - 1}  \biggr]^{-1}
$$
where
$$
t \geq \frac{\theta_3 + \sqrt{\theta^2_3 + 4}}{2} \phantom{abcde} \text{and} \phantom{abcde} \theta_j = \frac{\E[X^j]}{\sigma^j}
$$

Since the distribution $\Phi$ normalised $\E[X] = 0$ and $\sigma^2 = \E[X^2] = 1$, then $\theta_j = \E[X^j]$.  Furthermore, the kurtosis $\kappa = \E[X^4]$ is specified for the distribution.  $\E[X^3]$ may be expressed in terms of $N$ and $a$, the latter being a function of $N$ and $\kappa$:

\begin{prop}
$$
\E(X^3) = \frac{-3}{N-1} a + \frac{(N+1)}{(N-1)^2} a^3
$$

where $a$ is given by Proposition \ref{FormulaFora}:
$$
a = a(N,\kappa) = \sqrt{ -\frac{N-1}{N+1} + \sqrt{ \biggl( \frac{N-1}{N+1} \biggr)^2 - G(N,\kappa)}}
$$

Furthermore, $\E(X^3) \sim -3 (\kappa - 1)^{1/4} N^{-3/4} + (\kappa - 1)^{3/4} N^{-1/4}$ for large $N$.\\
\end{prop}

{\bf Proof:}

\begin{align*}
\E(X^3) = \sum_{i=1}^N X_i^3 & = \frac{a^3}{N} + \frac{N-1}{2N} (-b - \frac{a}{N-1})^3 + \frac{N-1}{2N} (b - \frac{a}{N-1})^3 \\
& = \frac{a^3}{N} - \frac{3b^2 a}{N} - \frac{a^3}{N(N-1)^2} \\
& = \frac{a^3 (N-1)^2 - 3 a \bigl( N(N-1) - a^2 N \bigr) - a^3}{N(N-1)^2} \\
& = \frac{-3}{N-1} a + \frac{(N-1)^2 + 3 N - 1}{N(N-1)^2} a^3\\
& = \frac{-3}{N-1} a + \frac{N(N+1)}{N(N-1)^2} a^3 \\
& = \frac{-3}{N-1} a + \frac{(N+1)}{(N-1)^2} a^3 \\
\end{align*}

For $N$ large, substituting $a = [(\kappa - 1) N]^{1/4}$ gives:

\begin{align*}
\E(X^3) & = \frac{-3}{N-1} [(\kappa - 1) N]^{1/4} + \frac{(N+1)}{(N-1)^2} [(\kappa - 1) N]^{3/4} \\
& \rightarrow -3 (\kappa - 1)^{1/4} N^{-3/4} + (\kappa - 1)^{3/4} N^{-1/4} \phantom{abcdefghij} \square
\end{align*}

Values of $\E(X^3)$ are tabulated below against $\text{Log}_{10}(N)$ for various values of kurtosis.  Interestingly, convergence to the limit of $\kappa - 1$ is not monotone, and requires a larger value of $N$ to converge than for $a$:\\

\begin{table}[h]
\begin{tabular}{| c || c | c | c | c |}
\cline{1-5}
& \multicolumn{4}{ | c | }{$\E(X^3)$} \\
\hline 		
$\text{Log}_{10}(N)$ &	 kurtosis = 7	&   kurtosis = 10 &	kurtosis = 13 &	kurtosis = 16 \\
\hline
 2  &  1.137  &  2.044  &  2.044  &  2.044 \\
 3  &  0.704  &  0.973  &  1.302  &  1.302 \\
 4  &  0.405  &  0.486  &  0.680  &  0.794 \\
 5  &  0.219  &  0.297  &  0.358  &  0.428 \\
 6  &  0.125  &  0.166  &  0.205  &  0.238 \\
 7  &  0.068  &  0.091  &  0.116  &  0.137 \\
 8  &  0.039  &  0.052  &  0.064  &  0.076 \\
 9  &  0.021  &  0.029  &  0.036  &  0.043 \\
 10  &  0.012  &  0.016  &  0.020  &  0.024 \\
\hline
\end{tabular}
\caption{\label{table5}Values of $\E(X^3)$.}
\end{table}


Setting $t = a$, values of Zelen's probability are given below:

\begin{table}[h]
\begin{tabular}{| c || c | c | c | c |}
\cline{1-5}
& \multicolumn{4}{ | c | }{$\biggl[ 1 + a^2 + \frac{(a^2 - a \theta_3 - 1)^2}{\theta_4 - \theta^2_3 - 1}  \biggr]^{-1}$} \\
\hline 		
$N-1$ &	 kurtosis = 7	&   kurtosis = 10 &	kurtosis = 13 &	kurtosis = 16 \\
\hline
 250  &  0.00398  &  0.00400  &  0.00396  &  0.00400  \\
 500  &  0.00199  &  0.00200  &  0.00200  &  0.00199  \\
 1000  &  0.00100  &  0.00100  &  0.00100  &  0.00100  \\
 10000  &  0.00010  &  0.00001  &  0.00010  &  0.00010  \\
 100000  &  0.00001  &  0.00001  &  0.00001  &  0.00001  \\
 1000000  &  0  &  0  &  0  &  0  \\
\hline
\end{tabular}
\caption{\label{table6}Values of Zelen's probability, $\biggl[ 1 + a^2 + \frac{(a^2 - a \theta_3 - 1)^2}{\theta_4 - \theta^2_3 - 1}  \biggr]^{-1}$ for the Bi-modal distribution with one extreme point.}
\end{table}

\medskip
\medskip

The inverse of the above probabilities are inverted to give ``1-in-'' values.  Thus, it can be seen that our distribution $\Phi_N$ gives a slightly tighter bound than Zelen, but only in the finite point case.

\begin{table}[h]
\begin{tabular}{| c || c | c | c | c |}
\cline{1-5}
& \multicolumn{4}{ | c | }{$1 + a^2 + \frac{(a^2 - a \theta_3 - 1)^2}{\theta_4 - \theta^2_3 - 1} $} \\
\hline 		
$N-1$ &	 kurtosis = 7	&   kurtosis = 10 &	kurtosis = 13 &	kurtosis = 16 \\
\hline
 250  &  251  &  250  &  253  &  250  \\
 500  &  503  &  501  &  500  &  502  \\
 1000  &  1000  &  1000  &  1002  &  1000  \\
 10000  &  10001  &  10001  &  10001  &  10001  \\
 100000  &  100000  &  100000  &  100000  &  100000  \\
 1000000  &  1000002  &  1000000  &  1000000  &  1000001  \\
\hline
\end{tabular}
\caption{\label{table1}Values of Zelen's probability for the Bi-modal distribution with one extreme point, given as ``1-in-'' values.}
\end{table}


\subsection{Bhattacharyya's Inequality}

Another extension was given by Bhattacharyya \cite{Ba}:
$$
P(X \geq t \sigma) \leq \frac{\kappa - \theta_3^2 - 1}{(\kappa - \theta_3^2 - 1)(1+t^2) + (t^2 - t \theta_3 - 1)}
$$
where
$$
t^2 - t \theta_3 - 1 > 0
$$

Analogous to Zelen's Inequality, set $\sigma = 1$ and $t = a$.  To satisfy $a^2 - a \theta_3 - 1 > 0$, $N$ must be suitably large, over 10,000 for our distribution $\Phi$.  Our distribution $\Phi_N$ gives a much more tighter bound than Bhattacharyya's inequality than it did for Zelen, but again, only in the finite point case.  For example, when $N = 10^7$ the probabilities are approximately 1-in-10,000.

\begin{table}[H]
\begin{tabular}{| c || c | c | c | c |}
\cline{1-5}
& \multicolumn{4}{ | c | }{$\frac{\kappa - \theta_3^2 - 1}{(\kappa - \theta_3^2 - 1)(1+a^2) + (a^2 - a \theta_3 - 1)}$} \\
\hline 		
$N-1$ &	 kurtosis = 7	&   kurtosis = 10 &	kurtosis = 13 &	kurtosis = 16 \\
\hline
 10,000  &  0.003453  &  0.002974  &  0.002646  &  0.002407  \\
 100,000  &  0.001099  &  0.000945  &  0.00084  &  0.000764  \\
 1,000,000  &  0.000349  &  0.000299  &  0.000266  &  0.000242  \\
 10,000,000  &  0.00011  &  0.000095  &  0.000084  &  0.000076  \\
 100,000,000  &  0.000035  &  0.00003  &  0.000027  &  0.000024  \\
\hline
\end{tabular}
\caption{\label{table8}Values of Bhattacharyya's probability, $\frac{\kappa - \theta_3^2 - 1}{(\kappa - \theta_3^2 - 1)(1+a^2) + (a^2 - a \theta_3 - 1)}$ for the Bi-modal distribution with one extreme point.}
\end{table}

\newpage

\section{Conclusion}

This paper considers how big a stress shock should be, based on how many multiples of the standard deviations are required to guarantee that the shock exceeds any observation within the history set.  By imposing the additional constraint of kurtosis, this lends itself to a more realistic description of asset price movements than that given by Samuelson's inequality \cite{S}.\\

A distribution $\Phi_N$ was constructed that maximises the number of standard deviations above the mean that a single point can be for a fixed value of kurtosis.  This was then used to validate several stress models, in particular that of Brace-Lauer-Rado \cite{BLR}, which has a primary parameter of kurtosis.  Furthermore, the constructed distribution can itself be used to derive stress shocks.\\

Following the outline of Samuelson's paper \cite{S}, we use our distribution to examine three extensions of Chebyshev's Inequality where the kurtosis is known.  It is found that our results give a tighter bound than the well-known inequalities, particularly that of  Bhattacharyya's Inequality \cite{Ba} is not, but only in the finite point case.

\section{Declarations of Interest}

The authors report no conflicts of interest. The authors alone are responsible for the content and writing of the paper.

\medskip
\medskip
\medskip
\medskip

\section{Appendix}

In Section 2, this paper uses the distribution of a bi-modal set of points, plus one outlier.  The intention is to have a distribution, which for a fixed value of kurtosis, enables the value of $a = (X_1 - \bar{X})/\sigma$ to be large.  As the bi-modal set of points has the lowest possible kurtosis, this should give the largest value of $a$ for any given kurtosis.  Indeed, at first glance Samuelson's example could be a candidate distribution that satisfies these requirements since, but the kurtosis here is $N$.\\

To rigorously construct a distribution, which for a fixed value of kurtosis, maximises the value of $(X_1 - \bar{X})/\sigma$, is a complex optimisation problem:

\begin{align*}
\max_{{X_i}} & \biggl\{ \frac{X_1 - \frac 1N \sum_{i=1}^N X_i}{\frac 1N \sum_{i=1}^N (X_i - \frac 1N \sum_{i=1}^N X_i)^2} \; | \\
& \phantom{abcde} \frac 1N \sum_{i=1}^N (X_i - \frac 1N \sum_{i=1}^N X_i)^4 / \bigl( \frac 1N \sum_{i=1}^N (X_i - \frac 1N \sum_{i=1}^N X_i)^2 \bigr)^2 = \kappa  \biggr\}
\end{align*}

To give some certainty this distribution cannot be improved upon, three other distributions were considered to test if the above construction could be bettered:\\

$\bullet$ A tri-modal distribution, with an equal number of points are $\{-1, 0, 1\}$ (kurtosis of 1.5),\\

$\bullet$ The distribution with two-thirds of the points at 0, and the remaining third at $1$ (kurtosis of 1.5),\\

$\bullet$ The Uniform distribution between $-1$ and $1$ (kurtosis of 1.8).\\

Since the statistic $(X_1 - \bar{X})/\sigma$ and the kurtosis are invariant under dilation and translation, the distributions are then re-scaled, and the outlier found $X_1 > 1$ using a simple search routine such that the kurtosis of the distribution is as desired.\\

Although each produced a value for $(X_1 - \bar{X})/\sigma$ close to the bi-modal, it was less in all cases:

\begin{table}[h]
\begin{tabular}{| c | c || c | c | c | c |}
\cline{1-6}
& & \multicolumn{4}{ | c | }{a = (Max-mean)/StDev} \\
\hline 		
N-1 &	sqrt(N-1)	& Bi-modal	&   Tri-modal &	Two-thirds &	Uniform \\
\hline
500 & 22.38 & 9.35 & 9.30 & 9.30 & 9.26 \\
1000 & 31.62 & 11.10 & 11.03 & 11.04 & 10.99 \\
2000 & 44.72 & 13.19 & 13.10 & 13.10 & 13.04 \\
3000 & 54.77 & 14.59 & 14.49 & 14.49 & 14.42 \\
4000 & 63.24 & 15.68 & 15.56 & 15.56 & 15.49 \\
5000 & 70.71 & 16.57 & 16.45 & 16.45 & 16.37 \\
10000 & 100 & 19.70 & 19.55 & 19.55 & 19.45 \\
\hline
\end{tabular}
\caption{\label{tableAppendix}Extreme point values for the considered example distributions.}
\end{table}

\medskip

Furthermore, as remarked in Section 2, $(X_1 - \bar{X})/\sigma$ does not differ very much for any choice of distribution.  This is desirable, and gives credibility to our choice of kurtosis being a good restriction to make, as it allows for asset prices to move - and move in different ways - with some impact on the volatility, not just from the one large shock.\\


\medskip
\medskip
\medskip




\begin{thebibliography}{12}

\bibitem{B}
Basel Committee on Banking Supervision (2018) Revisions to the minimum capital requirements for market risk, https://www.bis.org/bcbs/publ/d436.htm

\bibitem{Ba}
Bhattacharyya, B. B. (1987). One-sided chebyshev inequality when the first four moments are known, Communications in Statistics – Theory and Methods. {\bf 16}(9):2789$–$2791.

\bibitem{BLR}
Brace, A., Lauer, M., and Rado, M. (2006) A Stylised Model for Extreme Shocks: Four Moments of the Apocalypse, UTS Research Paper, https://www.researchgate.net/publication/23697087\_A\_Stylised\_Model\_for\_Extreme\_Shocks\_Four\_Moments\_of\_the\_Apocalypse

\bibitem{J}
Jensen, S. T. (1999) The Laguerre$–$Samuelson Inequality with Extensions and Applications in Statistics and Matrix Theory, MSc Thesis, Department of Mathematics and Statistics, McGill University.

\bibitem{M}
Maher, D. (2011) On the use of t-copulas for economic capital calculations, Journal of Risk Model Validation, {\bf 5}(3), 21-36.

\bibitem{MU}
 Mitzenmacher, M. and Upfal, E (2005). Probability and Computing: Randomized Algorithms and Probabilistic Analysis. Cambridge Univ. Press.

\bibitem{PR}
Platen, E., and Renata, R. (2008) Empirical Evidence on Student$-$t Log-Returns of Diversified World Stock Indices, Journal of Statistical Theory and Practice, {\bf 2}(2), 233-251.

\bibitem{S}
Samuelson, P. A. (1968) How deviant can you be?, J. Amer. Statist. Assoc., {\bf 63}, 1522-1525.

\bibitem{Sc} Scott, L. (1987) Option Pricing When the Variance Changes Randomly –
Theory, Estimation, and an Application, Journal Of Financial And
Quantitative Analysis, {\bf 22}(4), 419-438.

\bibitem{W} Wiggins, J. (1987) Option Values Under Stochastic Volatility – Theory
and Empirical Estimates, Journal Of Financial Economics, {\bf 19}, 351-372.

\bibitem{Z}
Zelen, M. (1954) Bounds on a distribution function that are functions of moments to order four, J Res Nat Bur Stand {\bf 53}, 377-381.



\end{thebibliography}
\end{document}